\documentclass{article}
\makeatletter

\@addtoreset{equation}{section}
\makeatother
\usepackage{epsf}
\begin{document}
\baselineskip 12pt
\title{Revisit to Low Mass Scalar Mesons  \\
 via Unitarized  Chiral Perturbation Theory}

\author{M. Uehara\thanks{e-mail: ueharam@cc.saga-u.ac.jp}\\
Takagise-Nishi 2-10-17, Saga 840-0921, Japan}
%\date{\today}
\maketitle
\begin{abstract}
We study how the scalar mesons below 1 GeV are generated through the Oller-
Oset-Pel\'aez version of the multichannel inverse amplitude method applied 
to the chiral perturbation theory. We find out that  the $f_0(980)$ state is 
generated as a bound state resonance below the $K\bar K$ threshold, 
while the $a_0(980)$ state, generated via the channel coupling between the 
$\pi\eta$ and $K\bar K$  channels, appears as a cusp at the $K\bar K$ threshold. 
The so-called $\sigma(500)$ and $\kappa(900)$ need not be interpreted as 
the conventional resonances.
\end{abstract}
%definitions%%%%%%
\def\beq{\begin{equation}}　\def\eeq{\end{equation}}
\def\beqa{\begin{eqnarray}}　\def\eeqa{\end{eqnarray}}
\def\beqan{\begin{eqnarray*}}　\def\eeqan{\end{eqnarray*}}
\def\ba{\begin{array}}　\def\ea{\end{array}}
\def\noeq{\nonumber}
\def\mpi{m_\pi} \def\fp{f_\pi}　\def\mK{m_K} \def\fK{f_K}
\def\me{m_\eta} \def\fe{f_\eta}　\def\half{\frac{1}{2}}　
\def\der{\partial}
\def\vg{\mbox{\boldmath$g$}}　\def\vt{\mbox{\boldmath$t$}}
\def\vG{\mbox{\boldmath$G$}}　\def\vK{\mbox{\boldmath$K$}}
\def\vS{\mbox{\boldmath$S$}}   \def\vT{\mbox{\boldmath$T$}}
\def\vrho{\mbox{\boldmath$\rho$}}
\def\del{\delta}
%%%%%%%%%%%%%%%%
\section{Introduction}
Recently, many papers have been appeared on the  studies of the 
scalar mesons below 1 GeV, but it seems that we have still not  reached 
any certain conclusion on their existence and nature as a whole.  

The experimental feature of the $(I,J)=(0,0)$ channel 
is qualitatively stated as follows: The phase shift  makes a plateau near 
$60^\circ\sim 80^\circ$, and reaches $90^\circ$ at $800\sim 900$ MeV 
and then rapidly increases to exceed  $200^\circ$ near the $K\bar K$ 
threshold,  and  $\pi\pi$ scattering is almost pure elastic below the $K\bar K$ 
threshold\cite{CERNMunich,Kaminski}. It is not clear, however, where  are 
the resonant states involving the so-called $\sigma$ meson  in the broad 
mass distribution from 400 to 1200 MeV denoted as $f_0(400\sim1200)$ in the 
Particle Data Group  Lists 2000\cite{PDG}. 
The complex pole searches in the $\pi\pi$ scattering amplitude, the 
attempts to fit the data by the Breit-Wigner formulae or the analyses 
by using the K-matrix  are cited there, but the obtained values 
are widely spread and seem to depend strongly on the models adopted. 
On the contrary the $f_0(980)$ state hidden in the structure near the $K\bar K$ 
threshold has the mass  converging  to a common value  near 980 MeV, while 
its width is spread a little\cite{PDG}.

In the $(I,J)=(1,0)$ channel  a sharp peak is observed near 
the $K\bar K$ threshold\cite{pietaMass}, which is denoted as $a_0(980)$. 
But it is still not clear  whether $a_0(980)$ is certainly a resonance or a 
cusp, since the $\pi\eta$ scattering data are scarce.  
About the $(I,J)=(1/2,0)$ channel, it is controversial whether the 
so-called $\kappa(900)$ really exists or not. The phase shift does not 
exceed $90^\circ$ below 1GeV at all\cite{piKdata}. This state is not yet cited 
in the PDG List\cite{PDG}. 

The central issue in the $(I,J)=(0,0)$ channel below 1 GeV is   
to understand the mechanism to generate the mass distribution such as 
the broad peak  and the dip or steep cliff near the $K\bar K$ threshold: Is
 the $\sigma$ state really hidden in the broad peak and  how the $f_0(980)$ 
state is generated ?
We are also interested in the following issues  whether 
the $a_0(980)$ state is also a bound state resonance like a twin of $f_0(980)$,
 and whether the $\pi K$ broad mass peak should be interpreted as the 
the resonance called by $\kappa(900)$.   \\

In this note we try to analyze the above issues by use of the multichannel 
Inverse Amplitude Method(IAM) of the Chiral Perturbation 
Theory\cite{DP, Hanna,GO,OOP,NP}.  In the IAM of the ChPT 
a meson-meson scattering amplitude is given as the Pad\'e\,[1,1] 
approximant applied to the perturbative amplitudes up to O($p^4$) in order to 
satisfy the $s-$channel unitarity and be applied to the resonance 
region.  The Lagrangian of the ChPT does not incorporate the $\sigma$ field 
nor other resonance fields as the independent degrees of freedom in advance, 
and then it offers a good theoretical framework to study the issues.  
Of course, this does not imply that the existence or  nonexistence of a resonance 
can be completely predicted by the IAM, since the theory contains a set of the  
 phenomenological parameters determined by the experimental data. 
 
Despite  the fact that the overall fits have already been given using the full 
$T_4$ amplitudes by Gom\'ez Nicola and Pel\`aez\cite{NP}, 
we revisit the issues around the scalar mesons with use of the 
IAM in order to understand the contents of the fits, that is,  how the resonances 
are generated or not generated. 
For this purpose we adopt the Oller-Oset-Pel\`aez (OOP) version of 
the IAM\cite{OOP}, since the OOP version is much simple and describes 
rather well the qualitative behavior of the data. 
 The OOP version picks up only the polynomial terms, $T^P_4$'s with 
 the coefficients $L_n$'s and the s-channel loop terms out of the 
 full O($p^4$) amplitudes,  neglecting the left hand singularities. 
 It may be said that the OOP version is like the K-matrix approximation 
 to the full amplitude.  \\

The conclusion on the scalar mesons viewed from the 
OOP version of the $2\times 2$ channel IAM is summarized as follows: 
\begin{enumerate}
\item The nonet structure of the  scalar mesons below 1 GeV does not hold.
\item The $f_0(980)$ state is a typical example 
of the bound state resonance\cite{DalitzTuan,FujiiU}: The $K\bar K$ scattering 
amplitude has a bound state pole on the real axis, 
if the channel coupling is switched off, and the channel coupling moves the pole 
into the second sheet and generates the resonant behavior near the 
$K\bar K$ threshold. But its explicit resonant form is hidden in the large 
$\pi\pi$  background. 
\item If we take the parameter set adopted in this note, the $a_0(980)$ 
state appears  as the strong cusp, not  as the resonance.  
The origin of $a_0(980)$ is not the $K\bar K$ bound state pole but the 
channel coupling between the $\pi\eta$ and $K\bar K$ channels. 
This  gives a sharp peak at the $K\bar K$ threshold, but the elastic 
$\eta\pi$ phase shift cannot exceed $90^\circ$ below the $K\bar K$ threshold. 
\item The broad peak of the isoscalar-scalar $\pi\pi$ mass distribution 
centered at 500 MeV need not be interpreted as the conventional resonance. 
It represents the  attractive interaction between two pions with the vacuum 
quantum number corresponding to the vacuum fluctuation coming from the 
spontaneous symmetry breaking. 
The $\kappa(900)$ state is also not the conventional resonance similar to 
$\sigma$.
\end{enumerate}
The mechanism to generate $f_0(980)$ and $a_0(980)$ is similar to the 
$K\bar K$  molecule model discussed by Weinstein and 
Isgur using the nonrelativistic quark model\cite{WeinIsg} 
and J\"ulich group using the meson exchange models\cite{Julich1,Julich2},  
though our theoretical framework is much different from theirs.\\

This paper is organized as follows: We 
briefly explain the Oller-Oset-Pel\`aez version  in the next 
section.  The scalar mesons are discussed in Sec.3 and 
the vector resonances  in Sec. 4. 
Concluding remarks and discussion are given in the last section. 

\section{The amplitudes in  the IAM }
Since the derivation of the IAM  applied to the ChPT  has 
been given in Refs.\cite{DP,Hanna,GO,NP}, here we summarize  the amplitudes in 
the OOP version\cite{OOP}. In this note we take the two-channel model.  
The partial wave amplitudes in the 
two-channel IAM to O($p^4$) are written as 
\beq
\vT(w)=\vt_2(w)[\vt_2(w)-\vt_4(w)]^{-1}\vt_2(w),
\eeq
where $\vt_2(\vt_4)$ is the partial wave amplitude with chiral 
order O($p^2$) (O($p^4$)) amplitude given by the symmetric $2\times2$ 
matrix  and 
$w$ is the total CM energy.  Our T-matrix is related to S-matrix as 
\beqa
S_{ij}(w)&=&\del_{ij}-2i\rho_i^{1/2}(w)T_{ij}(w)\rho_j^{1/2}(w),\\
\rho_i(w)&=&\frac{1}{16\pi}\frac{2k_i}{w}
\eeqa
where $\rho_i$ is the phase space 
factor with $k_i$ being the CM momentum in the i-th channel.

Instead of taking the full $T_4(s,t,u)$, which involves  
the left-hand cut coming from the $t-$ and $u-$channel 
loop diagrams, the OOP version picks up only the s-channel loop diagrams 
and the polynomial  terms written as $T^P_4(s,t,u)$ with the phenomenological 
coefficients $L_n$'s left after the renormalization of the one-loop 
diagrams\cite{GL2,Bernard}. 
All of our $T^P_4(s,t,u)$ are taken from the Ref. \cite{NP}.
Thus, the partial wave $\vt_4(w)$ is given as
\beq
\vt_4(w)=\vt^P_4(w)+\vt_2(w)\cdot\vG(w)\cdot\vt_2(w), \label{OOPamp}
\eeq
where $\vt^P_4(w)$ is the partial waves coming from $T^P_4(s,t,u)$, and 
$\vG(w)$ is the diagonal $2\times 2$ loop integral under the 
$\overline{MS}-1$ regularization \cite{GL2}. $G_i(w)$ in the $i-$th 
channel is written as 
\beqa
G_i(w)&=&\frac{1}{(4\pi)^2}\left\{-1+\log\left(\frac{m^2}{\mu^2}\right)+
\sigma_i(w)\log\left(\frac{\sigma_i(w)+1}
{\sigma_i(w)-1}\right)\right\},\\
\sigma_i(w)&=&\sqrt{1-4m^2/w^2}=\frac{2k_i}{w}
\eeqa
for the channel with the equal mass $m$ in the $i-$th channel, and $k_i$ is the CM 
momentum, and 
\beqa
G_i(w)&=&\frac{1}{(4\pi)^2}\left\{-1+
\log\left(\frac{m_1m_2}{\mu^2}\right)
+\frac{\Delta_{21}}{w^2}\log\left(\frac{m_2}{m_1}\right)\right.
 \noeq\\
&+&\left. \lambda_i(w)\log\left(
\frac{\sigma_{i+}(w)+\sigma_{i-}(w)}
{\sigma_{i+}(w)-\sigma_{i-}(w)}\right)\right\},\\
\sigma_{i\pm}&=&\sqrt{1-(m_1\pm m_2)^2/w^2},\\
\lambda_i(w)&=&\sigma_{i+}(w)\sigma_{i-}(w)
\eeqa
for the channel with unequal masses $m_1$ and $m_2$ in the $i-$th channel, and 
$\Delta_{21}=m_2^2-m_1^2$. The imaginary part of $G_ii(w)$ is 
given as 
\beq
{\rm Im}G_i(w)=-\frac{1}{16\pi}\frac{2k_ii}{w}\theta(w-w_i)
\equiv -\rho_i(w)
\eeq
with $w_i$ is the threshold energy of the $i-$th channel. 

Although the full amplitudes $T_2$ and $T_4$ satisfy the crossing symmetry, 
it is known that the unitarized amplitudes projected on the definite 
$(I,J)$ channel of the $s-$channel break the crossing symmetry.  
The amplitudes of the OOP version could break 
 it more largely, since the version discards the $t-$ and 
$u-$channel loop diagrams. \\

The OOP version of the multichannel IAM is just like the K-matrix 
formalism, which ignores the lefthand cut of the scattering amplitudes. 
Defining the K-matrix as 
\beq
\vK=\vt_2[\vt_2-{\rm Re}(\vt_4(w))]^{-1}\vt_2, \label{Kmatrix}
\eeq
where $\vt_4$ is the OOP version given by Eq.(\ref{OOPamp}), 
we rewrite the OOP version of the IAM amplitude as 
\beq
\vT=\vK[1+i\vrho\cdot\vK]^{-1}. 
\eeq
$\vK$ in Eq.({\ref{Kmatrix}) is real analytic as the usual K-matrix.

Since the OOP $\vt_4$ discards the contributions from the left hand cut,  
the results  would be  different from those obtained by using the full $T_4$ 
amplitudes under the same set of the coefficients $L_n$'s\cite{BP}.  
Thus, we relax the values of $L_n$'s obtained in ref.\cite{NP} 
in order to reproduce  the characteristic features of the scattering amplitudes.   
We adopt the parameter set of $\hat L_n\times 10^3$ shown in 
Table\,\ref{tab:Ln}, which is compared  with the set of ref.\cite{NP} cited as 
GNP and the one-loop ChPT set obtained in 1995\cite{Ecker}.  Note that 
while the GNP and the ChPT sets are determined at $\mu=M_\rho$, 
we take  $\mu=1$ GeV. This is 
because  a little larger $\mu$  is easy to reproduce the qualitative 
behavior of the $(I,J)=(0,0)$ channel. The original OOP version took  the 
momentum cutoff corresponding to $\mu=1.2$ GeV\cite{OOP}. For the vector 
meson channel much larger $\mu$ can give better results.  

\def\s{\small}
\begin{table}[h]
\label{tab:Ln}
\begin{center}
\begin{tabular}{|c|r|r|r|r|r|r|r|r|}\hline
&$\hat L_1$&$\hat L_2$&$\hat L_3$&$\hat L_4$&$\hat L_5$&$\hat L_6$
&$\hat L_7$&$\hat L_8$ \\ \hline
GNP&0.56&1.21&$-2.79$&$-0.36$&1.4&0.07&$-0.44$&0.78\\ \hline
Ours &0.51&1.11&$-2.95$&$-0.35$&1.5&$-0.2$&$-0.4$&0.89\\ \hline
{\s ChPT}&{\s $0.4\pm 0.3$}&{\s $1.35\pm 0.3$}&{\s $-3.5\pm 1.1$}&{\s $-0.3\pm 0.5$}
&{\s $1.4\pm 0.5$}&{\s $-0.2\pm 0.3$}&{\s $-0.4\pm 0.2$}&{\s $0.9\pm0.3$}\\ \hline
\end{tabular}
\caption{$\hat L_n\times 10^3$}
\end{center}
\end{table}
It should be stressed that our present set is not the best choice to fit 
the experimental data: Indeed, 
we fail to reproduce the isospinor $\pi K$ scattering amplitude  below 770 MeV. 
It is interesting, however, that our set  is not far from the GNP and ChPT 
sets.\footnote{Other sets of $\hat L_n$'s in refs.\cite{OOP,GO} are not quoted, 
because their amplitudes are different from those of ref.\cite{NP}.}

Now, we summarize the notations used in the following sections. 
We divide  $D_{ij}=[\vt_2-\vt_4]_{ij}$ into the real and imaginary part as
\beq
D_{ij}(w)=a_{ij}+i(\rho_1 b^1_{ij}+\rho_2 b^2_{ij}),
\eeq
where
\beqa
a_{ij}&=&(\vt_2)_{ij}-(\vt^P_4)_{ij}-(\vt_2\vg(w)\vt_2)_{ij},\label{aij}\\
b^k_{ij}&=&(\vt_2)_{ik}(\vt_2)_{kj}\label{bijk}
\eeqa
with  $\vg(w)={\rm Re}\vG(w)$.  Thus, $T_{ii}(w)$, the elastic scattering 
amplitude of the i-th channel, is written as 
\beq
\rho_iT_{ii}(w)=\frac{B_i(w)+iC(w)}{A(w)
+i[B_1(w)+B_2(w)]},\label{Tij}
\eeq
where
\beqa
A(w)&=&a_{11}a_{22}-a_{12}^2-C(w), \label{A}\\
B_1(w)&=&\rho_1[a_{11}b^1_{22}+a_{22}b^1_{11}-2a_{12}b^1_{12}],
\label{B1}\\
B_2(w)&=&\rho_2[a_{11}b^2_{22}+a_{22}b^2_{11}-2a_{12}b^2_{12}],
\label{B2}\\
C(w)&=&\rho_1\rho_2[b^1_{11}b^2_{22}+b^1_{22}b^2_{11}-
b^1_{12}b^2_{12}]. \label{C}
\eeqa
The $i-$th channel amplitude $\hat T_{ii}$ in the single channel problem 
is given as
\beq
\rho_i\hat T_{ii}(w)=\frac{\rho_ib^i_{ii}}{a_{ii}+
i\rho_i b^i_{ii}+i\rho_jb^j_{ii}}, \label{hatTii}
\eeq
where we note that $a_{ii}$ contains  the meson loops not only of the 
$i-$th  channel but also of the other $j-$th channel. 

The S-matrix for the first channel with the lower threshold energy is 
written as 
\beq
S_{11}=\frac{A+2C-i(B_1-B_2)}{A+i(B_1+B_2)}=\eta(w)\exp[2i\del_1(w)], 
\label{S11}
\eeq
where the  phase shift $\del_1(w)$ 
and the inelasticity $\eta(w)$ are calculated through this definition.

\section{Scalar mesons}
We study how the scalar mesons are generated through the IAM applied to 
ChPT. Since we need qualitative behavior of the amplitudes, 
we do not cite any experimental data to compare the calculated results. 
The definitions of the amplitudes have been given in Eqs.(\ref{aij}), 
(\ref{bijk}), (\ref{A}), (\ref{C}), (\ref{B1}),(\ref{B2}). 
We calculate up to 1.2 GeV within the OOP version of the two-channel 
IAM, but we should note that the $\eta\eta$ channel and higher resonances 
may affect the results even below 1.2 GeV. 
 
 \subsection{The (I,J)=(0,0) channel: $\pi\pi\times K\bar K$}
 The characteristic behavior of this channel is that the phase shift 
 $\del_{00}(w)$ rises from the $\pi\pi$ threshold, forms a plateau of 
 $60^\circ\sim 80^\circ$  from 500 to 800 MeV, 
 crosses $90^\circ$ in the region $800\sim 900$ MeV, and suddenly increases 
 to exceed $200^\circ$just below the $K\bar K$ threshold. 
 \begin{figure}[h!]
\begin{center}
 \epsfxsize=12cm
 \centerline{\epsfbox{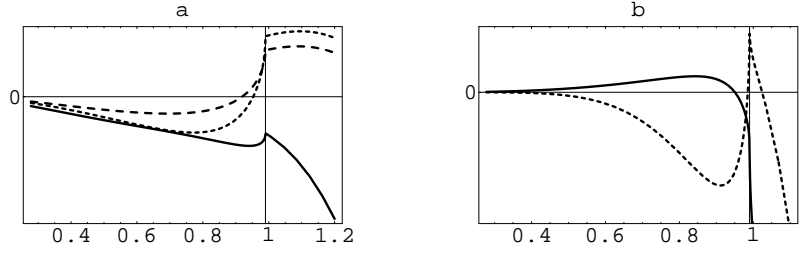}}
 \end{center}
 \label{fig:amp00}
 \caption{Energy dependence of $a_{ij}(w)$. (a) $a_{11}$ is given by 
 the solid line, $a_{22}$ by the dotted line and $a_{12}$ by the dashed line. 
 (b) Energy dependence of the resultant amplitude. The real part, $A$, 
  is given by the solid line and the imaginary part,$B_1+B_2$, by the 
  dotted line. }
 \end{figure}
The behavior implies that $A(w)$ defined by Eq.(\ref{A}),  
 the real part of denominator of $T_{11}(w)$, 
 should develop a zero at the point where $\delta_{00}$ goes across
 $90^\circ$ and $B_1(w)$ of 
 Eq.(\ref{B1}), the imaginary part, does so at very close to the $K\bar K$  
 threshold. 
 \begin{figure}[h!]
 \begin{center}
 \epsfxsize=12cm
 \centerline{\epsfbox{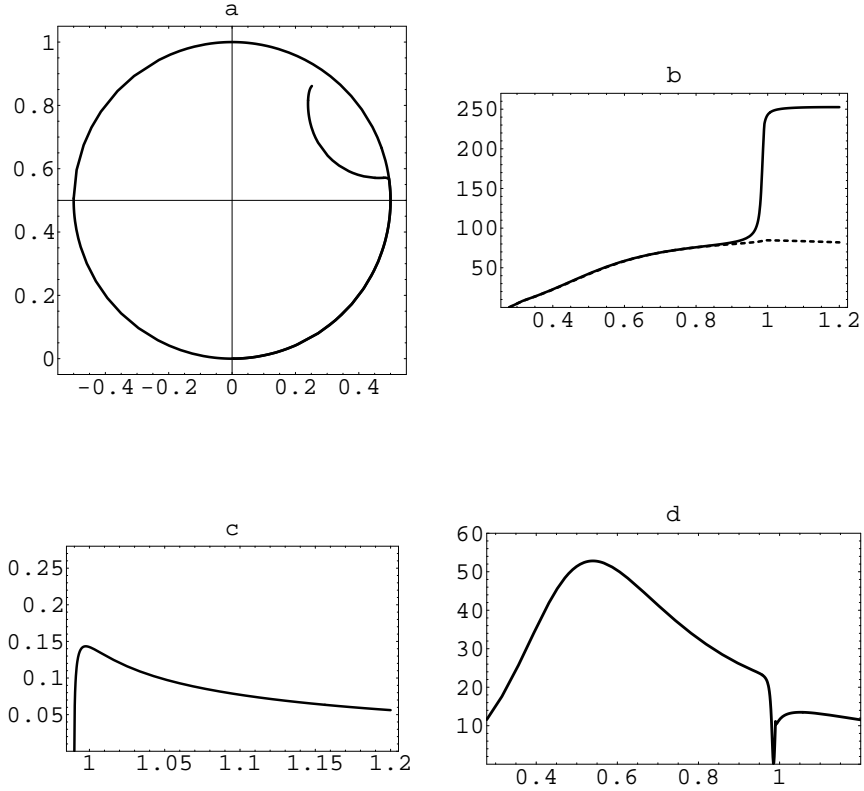}}
 \label{fig:fig00}
 \caption{(a) Argand diagram starting at the $\pi\pi$ threshold, drawing 
 the circle and bending at $K\bar K$ threshold.. (b) Phase shift $\del_{00}$
 given by solid line, and the single channel  phase shift by dotted line.
 (c) Absorption rate, $(1-\eta_{00}^2)/4$. (d) The $\pi\pi$ cross section in 
 mb.}
 \end{center}
 \end{figure}

 At first we observe from Fig.1(a) that $a_{22}(w)$ develops a 
 zero. This zero stays below the $K\bar K$ threshold,  if we eliminate 
the pion-loop contribution in $a_{22}(w)$. This means that there is the bound 
state pole in the isolated $K\bar K$ channel, which generates the bound state 
resonance in the $\pi\pi$ scattering amplitude with $(I,J)=(0,0)$ state. 
The bound state pole moves into the second  
complex sheet near the real axis as the channel coupling is switched on.  The 
nearby pole appears as the zero of $A(w)$. The zero of $B_1(w)$ is also due to the 
 zero of $a_{22}\cdot b^1_{11}$ primarily, but the remaining negative 
 terms push the zero closer to the $K\bar K$ threshold. It is noted, however, 
 that the negative terms are required to be not so large in order 
 to maintain the zero in $B_1$. If $B_1$ does not develop the zero, 
 the resultant phase shift cannot exceed $180^\circ$. Thus, 
 the sudden rise of the phase shift is not the simple result of the bound state 
 pole in the $K\bar K$ channel. 
  This mechanism to generate the bound state resonance below $K\bar K$  
threshold  is similar to  the one discussed by Weinstein and Isgur\cite{WeinIsg} 
 and J\"ulich group\cite{Julich1, Julich2}, though the theoretical 
 framework of the IAM is quite different from  theirs. 
 
We show the results on the Argand diagram, the phase shift $\del_{00}$, 
the absorption rate $[1-\eta_{00}^2]/4$ and the $\pi\pi$ cross section 
in Fig.2 (a), (b) and (c), respectively. 
 Our calculation reproduces the characteristic features qualitatively, 
 and  gives the results very similar to those by the full amplitude 
calculation\cite{NP}. 
\begin{center}
 \begin{figure}[h!]
 \epsfxsize=12cm
 \centerline{\epsfbox{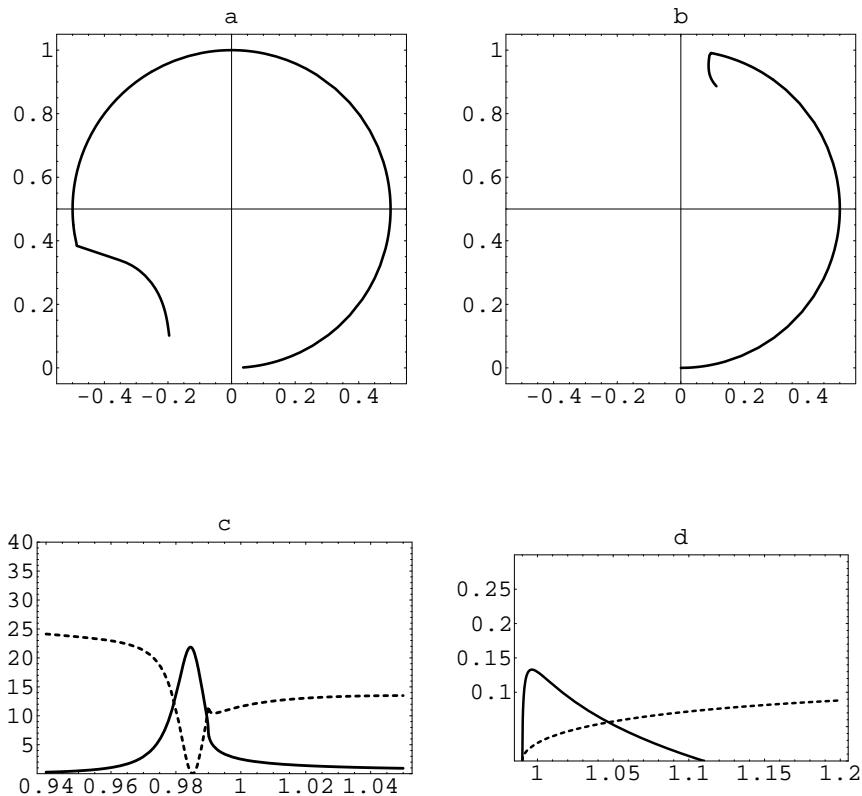}}
 \label{fig:f0amp}
 \caption{(a) The Argand diagram of $f_0$ amplitude 
 from 900 to 1050 MeV. (b) The Argand diagrams of $\hat T_{11}$ from the 
 $\pi\pi$ threshold to 1.2 GeV. 
 (c) The solid line  shows the sharp Breit-Wigner peak of the  
 the $f_0$ state extracted by Eq.(\ref{f0}), dotted line does the full 
 cross section, and (d) The absorption rate $(1-\eta_{00})^2)/4$ of 
 $f_0(w)$ is given by the solid line and the one of  
 the "background"  by the dotted line. }
 \end{figure}
 \end{center}

Now, in order to extract the $f_0(980)$  behavior explicitly we define the $f_0$ 
amplitude through  the following formula;
\beq
T_{11}(w)=\hat T_{11}(w)+f_0(w)\exp[2i\hat\del_{11}(w)], \label{f0}
\eeq
which is the Dalitz-Tuan prescription widely used to unitarize the sum of 
the two scattering amplitude\cite{DalitzTuan}. 
 Here  $\hat T_{11}$ is the single channel $\pi\pi$  amplitude with 
$\hat S_{11}=1-2i\hat T_{11}$ defined in Eq.(\ref{hatTii}).  $\hat T_{11}$ 
cannot draw a full circle in the energy range to 1.2 GeV and plays a role of the 
background to $f_0(w)$.  We observe that  a sharp  peak of the $f_0$ state 
hidden in the $\pi\pi$ cross section appears like the typical Breit-Wigner 
resonance as shown in Fig.3 (a) and (c), and that the strong rise of the 
absorption rate $(1-\eta_{00}^2)/4$ just above the $K\bar K$  threshold should be 
attributed to the resonant behavior of the $f_0$ amplitude as shown 
in Fig.3 (d). The $f_0$ amplitude could be approximated by the Breit-Wigner 
form,
\beq
f_0(w)=\frac{m_0\Gamma_1}{m_0^2-s+im_0(\Gamma_1+\Gamma_2)} \quad\mbox{ with }\Gamma_i=g_i^2w\rho_i,
\eeq
where $m_0$ is the mass of $f_0$, $g_1(g_2)$ the scalar coupling constant 
 to $\pi\pi\,(K\bar K)$. In order to reproduce the sharp rise of the absorption 
rate just above the $K\bar K$ threshold, the ratio $g_2^2/g_1^2$  is required 
to be large as the mass $m_0$ leaves the $K\bar K$ threshold. We find that 
$m_0\sim 985$ MeV, $g_1^2\sim 0.5$ and $g_2^2/g_1^2\sim 10$ in our set 
of the parameters, but the detailed values of the mass 
and  width are too model-dependent and cannot be trusted. 
 At least it would be certain that 
$f_0$ behaves like the Breit-Wigner resonance with the large $K\bar K$   
coupling constant.

The rising and plateau of the phase shift  and the falling behavior of the 
kinematical limit of the cross section, $4\pi/k^2$, shape the $\pi\pi$ cross 
section into the broad peak from $400\sim 800$ MeV, peaked at 
500 MeV, as shown in Fig.2 (d).  We show also 
 that the phase shift below 800 MeV is almost similar to that of 
 the single channel $\hat T_{11}$, and that $\hat T_{11}$ cannot draw 
 the full Argand circle up to 1.2 GeV. This behavior is generated essentially by 
chiral symmetry and unitarity\cite{CGL}, and  we do not need  any mechanism 
to hide a conventional resonance in the broad peak.  
Such a strong two-pion correlation may be called 
the $\sigma$ state, but it needs not be expressed by an unstable particle as 
in the linear $\sigma$ model, or by a Breit-Wigner formula. 

\subsection{The (I,J)=(1,0) channel: $\pi\eta\times K\bar K$}
 The $a_{22}$ amplitude  has a zero similar to the (0,0) channel, but we note 
that there are large differences; the amplitude $a_{11}$  is weak repulsive, 
the $\pi\eta\to (K\bar K)_{I=1}$  amplitude $a_{12}$ is large, and 
the zero has disappeared in the resultant amplitude $A$ contrary to 
the (0,0) channel.  Where does the zero disappear ? In order to see the role of 
the channel coupling including the $\pi\eta$ loop in $a_{22}$ and $K\bar K$ 
loop in $a_{11}$, we introduce a scale factor $\alpha$ as $\alpha(t_2)_{12}$ and 
$\alpha(t_4)_{12}$. We observe that the zero of $a_{22}(w)$ does not develop 
at $\alpha=0$,  that is, there is not a bound state pole. When increasing 
$\alpha$ gradually, the cusp behavior becomes remarkable and then the cusp 
changes to the resonance for a sufficiently large $\alpha$. 
Such a behavior was studied using a much simpler model in  \cite{FujiiU}, where 
the complex pole pole starts from the boundary between the third  and 
fourth sheets at $\alpha=0$, moves to the fourth sheet producing the cusp 
behavior as $\alpha$ increases, and then appears in the second sheet as 
the resonance pole. 
 At $\alpha=1$ in our $(1,0)$ channel, the pole stays in the fourth sheet near the 
boundary of the second sheet , and then the strong cusp behavior is left. The real 
part of the complex pole is larger than $2m_K$. 
The cusp gives a sharp peak of 
the $\pi\eta$ cross section, the shape of which is not expressed by the 
Breit-Wigner form as shown in  Fig. 5(d). 
\begin{figure}[h!]
 \begin{center}
 \epsfxsize=12cm
 \centerline{\epsfbox{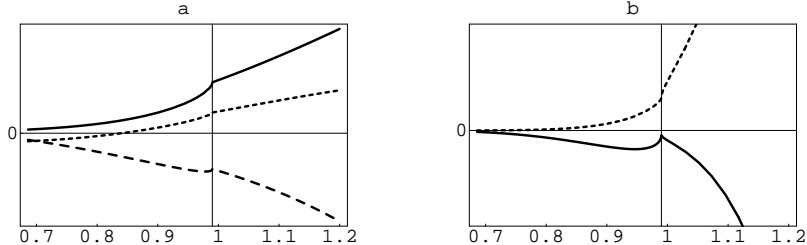}}
 \label{fig:amp10}
 \caption{Energy dependence of the amplitudes. (a) $a_{11}$ is given by 
 the solid line, $a_{22}$ by the dotted line and $a_{12}$ by the dashed line. 
 (b) $A$ is given by the solid line and $B_1+B_2$ by the dotted line. }
 \end{center}
 \end{figure}
 \begin{center}
 \begin{figure}[h!]
 \epsfxsize=12cm
 \centerline{\epsfbox{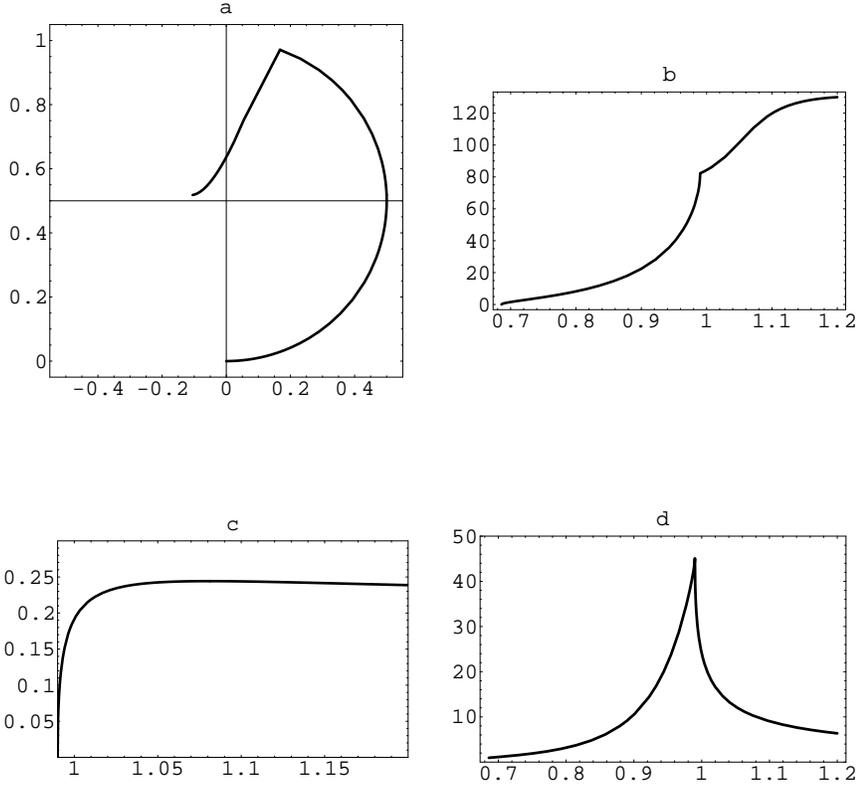}}
 \label{fig:fig10}
 \caption{(a) Argand diagram. (b) Phase shift $\del_{10}$.
 (c) The production rate. (d) The $\pi\eta$ cross section in units of 
 mb. }
 \end{figure}
 \end{center}

The large $\pi\eta\to (K\bar K)_{I=1}$ amplitude drags $T_{11}$ 
toward the center in the Argand diagram  above the $K\bar K$ threshold 
 as shown in Fig.5 (a), and  the phase shift increases 
 to pass $90^\circ$ {\em above} the $K\bar K$ threshold. The behavior  of 
 the phase shift above the $K\bar K$ threshold does not imply the 
 existence of a resonance above the $K\bar K$ threshold, however.   
 It is also noted  that the form of the absorption rate $(1-\eta^2_{10})/4$ 
 is quite different from the one of the resonant $f_0(980)$ state which 
 shows the sharp rising  just above the $K\bar K$ threshold. Although 
it is not clear that the full $T_4$ calculation reached the same cusp behavior 
because of the lack of the figure of the phase shift, we point out that 
the cross section behaves very similarly to our result\cite{NP}.\\

 It is interesting to  note that our result resembles the J\"ulich
model\cite{Julich2}, where  they state that the  $a_0$ state is essentially 
generated by the channel coupling without any origin in the diagonal amplitudes, 
both $\pi\eta$ and $K\bar K$ channels. They seem to choose finally the 
parameters so that the pole is brought to the second sheet off 
the fourth sheet and then the round shaped $\pi\eta$ cross section 
is obtained. 
 
\subsection{$(I,J,)=(1/2,0)$ channel: $\pi K\times\eta K$}
The experimental $\pi K$ phase shift increases smoothly up to 
the $\eta K$ threshold without any remarkable structure, 
but does not exceed $90^\circ$\cite{piKdata}. This 
behavior implies that there is not hidden any conventional resonant structure. 

\begin{figure}[h!]
 \begin{center}
  \epsfxsize=12cm
 \centerline{\epsfbox{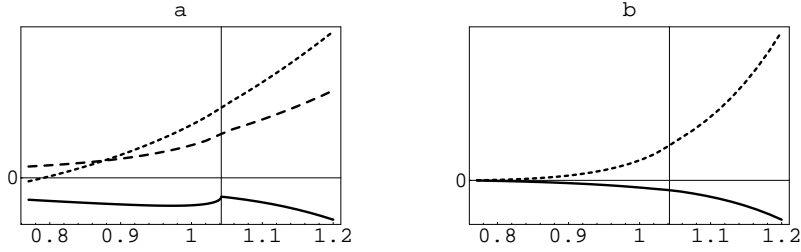}}
 \label{fig:amp1/20}
 \caption{Energy dependence of the amplitudes. Lines are the same as previous 
Figs. }
\end{center}
\end{figure}
 \begin{figure}[h!]
 \begin{center}
 \epsfxsize=12cm
 \centerline{\epsfbox{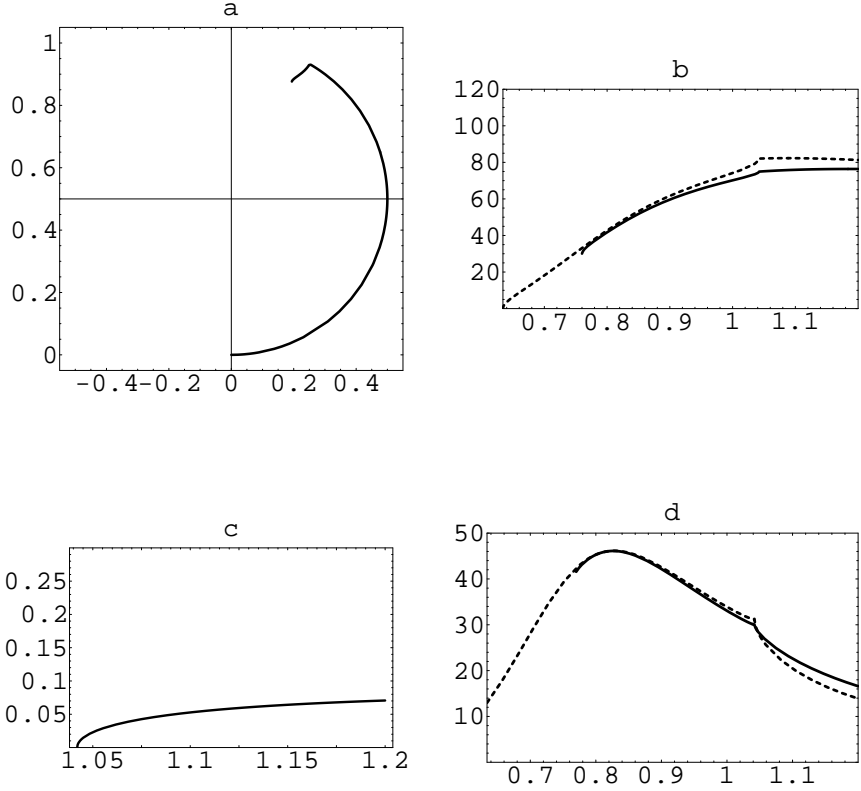}}
 \label{fig:fig1/20}
 \caption{(a) Argand diagram. (b) Phase shift $\del_{\frac{1}{2} 0}$ 
 The solid  line is the multichannel calculation, and the dotted one 
 for the single  channel calculation.
 (c) The production rate. (d) The $\pi K$ cross section. }
 \end{center}
  \end{figure}
Due to the wrong sign of the $\eta K$ elastic amplitude $r_{22}$ 
at low energies below 760 MeV,  both of the $A$ and $B_1$ have the wrong 
sign, but we can reproduce the above experimental behavior roughly above 
770 MeV.  The failure is directly due to the fact that 
the leading $\eta K\to\eta K$ amplitude $(t_2)_{22}$  behaves as if 
it has a bound state zero at a low energy and $t_4$ cannot remove the zero. 
This would be  due to the OOP version with our set of parameters  
$L_n$'s, since the calculation by the full amplitudes gives the nice 
result\cite{NP}. The Argand diagram shows that the amplitude stops in the 
second quadrant at $\eta K$ threshold, and the phase shift increases up to about 
$80^\circ$,  which is a little larger than the data, but does not exceed $90^\circ$ 
below 1.2 GeV. The Argand diagram up to 1.2 GeV is similar to the result in 
Ref.\cite{WeinIsg2}. 
The phase shift by the single $\pi K$ amplitude is very similar to 
the multichannel one at higher energies as shown in Fig.7(b). This fact and the 
rather small absorption by the $\eta K$ channel indicate that the effects of 
the  channel coupling seems to be small as contrasted with  the $f_0(980)$ 
and $a_0(980)$ cases. The $\pi K$ cross section has the large broad peak 
centered at 850 MeV.

It is certain that both the calculation and experimental data in the (1/2,0) 
channel lead to the conclusion that that 
the so-called $\kappa$ represents the strong $\pi K$ correlation, but not 
the conventional resonance. The absence of this 
state  is  argued in  Refs.\cite{Cherry,Torn,WeinIsg2}.\\

\subsection{$(I,J)=(2,0)$ and $(3/2,0)$ exotic channels}
The $\pi\pi_{(2,0)}$ channel is repulsive, and the phase shift goes 
 down as shown in Fig.\ref{fig:exotic}(a). Similarly, the $\pi K$ channel 
 gives the negative $(3/2,0)$ phase shift, which is shown in 
 Fig.\ref{fig:exotic}(b). These behaviors are very similar to those of 
 ref.\cite{NP}, and to the experimental data.

\begin{figure}[h!]
\begin{center}
\epsfxsize=12cm
\centerline{\epsfbox{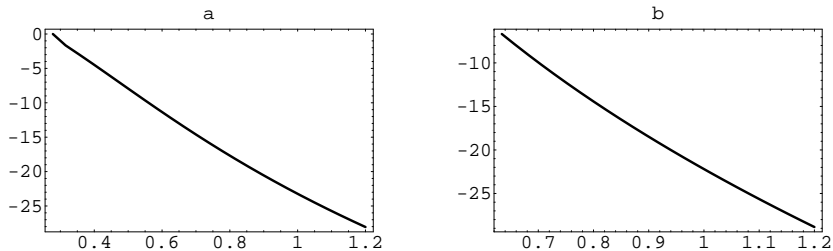}}
 \label{fig:exotic}
 \caption{(a) $\del_{(2,0)}$. (b) $\del_{(3/2,0)}$.}
 \end{center}
 \end{figure}

\subsection{Scattering lengths}
Here we summarize the scattering lengths of $(I,J)=(0,0)$, $(2,0)$,  
$(1/2,0)$ and $(3/2,0)$ channels in unit of $1/m_{\pi}$: 
\beqan
a_{(00)}=0.214, &\qquad& a_{(10)}=0. 029   \qquad a_{(1/2,0)}=/,\\
a_{(2,0)}=-0.040, &\qquad& a_{(3/2,0)}=-0.057.
\eeqan
While the value $a_{(1/2,0)}$ cannot be trusted by the situation stated in 
the previous subsection,  the  $\pi K$ single channel calculation gives a good  
value, $a_{(1/2,0)}=0.219$.  The channel coupling also changes the scattering 
length given by the single $\pi\eta$ channel calculation in the $(1,0)$ channel.
The scattering lengths by the one-loop ChPT to $\pi\pi$ 
scattering\cite{a00} and $\pi K$ scattering under the condition
 $f_\pi=f_K$\cite{Bernard} are as follows:
\beqan
a_{(00)}=0.201 &\qquad & a_{(2,0)}=-0.041,\\
a_{(1/2,0)}=0.19 &\qquad & a_{(3/2,0)}=-0.05,
\eeqan
 which are close to our calculations, if we take the single channel 
result for $a_{(1/2,0)}$. 

\section{Vector mesons}
We do not attempt to reproduce the resonance masses exactly, but 
we think that it is enough for our aim to reproduce them  within  the 
error of a few tens MeV . 

We define the kinematical singularity free amplitudes
$t_{ij}(w)$ as 
\beq
T_{ij}(w)= k_i\cdot t_{ij}(w)\cdot k_j,
\eeq
where $k_i$ is the CM momentum in the $i-$th channel. This replacement 
induces the extra momentum dependence on $\vG$ as 
\beq
\vG_i(w)\quad \rightarrow\quad G_i^P(w)=k_i^2G_i(w). 
\eeq
Then, the phase space factor $\rho$ becomes 
\beq
\rho^P_i(w)=k^2_i\rho_i(w). 
\eeq

At first, we observe that the vector meson resonances are realized 
even in the single channel formalism, and that the channel coupling do 
not affect the result so much, if the parameter set $\hat L_n$'s is 
selected appropriately. We tabulate the results at $\mu=1$ GeV in 
Table \ref{tab:Vres}.
\begin{table}[h]
\label{tab:Vres}
\begin{center}
\renewcommand{\arraystretch}{1.2}
\begin{tabular}{|c|c|r|r|r|}\hline
& &single&multi& Exp. value\\ \hline
$\rho$ &Mass(MeV)&796.9&795.6&$769.3\pm 0.8$ \\ \cline{2-5}
&Width(MeV) & 165.8&165.2 &$150.2\pm 0.8 $\\ \hline
$K^*$&Mass(MeV)& 885.2&859.9&$891.7\pm 0.3$ \\ \cline{2-5}
&Width(MeV)& 40.1&36.5& $50.8\pm 0.9$ \\ \hline
"$\phi$"&Mass(MeV)& 887.9& &926.5 \\ \hline
\end{tabular}
\caption{Masses and widths of the vector resonances both in 
the single and multichannel formalism.}
\end{center}
\end{table} 
As to the $phi(1020)$ meson  Ref.\cite{NP} gives 935 MeV by use of the full 
$T_4$, and they argue that this state  is  
the SU(3) octet part of the  $\phi$ meson, because $P-$wave $K\bar K$ 
state can couple only to 
the SU(3) octet state, the mass of which is given as 926.5 MeV by 
using the experimental masses of the vector mesons\cite{ChengLi}. If 
we take this value as the experimental mass of "$\phi$", the 
deviation of our results from the experimental mass values of the vector 
meson remains within $\pm 40$ MeV.  Since the value of the width 
depends almost linearly to the mass value, the calculated 
width is large by about 10\% for the $\rho$ meson and small by 30\% 
for the $K^*$ meson compared to the experimental one, respectively. 
If we use the experimental mass values,  
the calculation gives the $\rho$ width  154.2 MeV and $K^*$ 45.8 MeV, 
the both of which are reasonable. \\

In the vector meson channels both for the $\rho$ and $K^*$, all of the 
$a_{ij}(w)$ amplitudes cross the zero at almost similar energy. 
Order of the positions of the zeros in the $rho$ channel is different 
from those in the $K^*$ channel. This is the reason why the mass in 
the multichannel is so close to the one in 
the singlechannel for the $\rho$ channel, but large for the $K^*$ channel.  
The subtle positions of the zeros depend on the parameter set 
$\hat L_n$'s, but the developing of the zeros in all of $a_{ij}$ would 
be stable. The developing of the common zeros is the remarkable  
difference from the scalar channels. 

\section{Concluding remarks}
We have calculated the two-meson scattering 
amplitudes, mainly in the scalar channel, within the OOP version 
of the IAM to the ChPT.  We used the $T^P_4$ amplitudes 
calculated by Gom\'es Nicola and Pel\`aes\cite{NP} with our set of the 
phenomenological constants $\hat L_n$'s given in Table I.  \\

We have arrived at the following conclusion on the low mass 
scalar meson spectroscopy.
\begin{enumerate}
\item The nonet structure of the scalar mesons below 1 GeV does not hold.
\item The $f_0(980)$  state is 
generated through the bound state appearing  in the $K\bar K$   
channel,  and behaves as like as the Breit-Wigner resonance, if we  
extract the $f_0$ amplitude from the $\pi\pi$ background.
\item The $a_0(890)$ state is born through the channel coupling between the 
$\pi\eta$ and $K\bar K$ channels, and finally grows up to be the strong cusp.  
\item The broad peak centered at 500 MeV, which may be called the 
$\sigma$ {\em state}, is interpreted as  the $\pi\pi$ strong correlation   
coming from chiral symmetry and unitarity,  and there is no conventional 
resonance below 900 MeV. 
\item Unfortunately, we failed to 
reproduce the low energy behavior of the $(I,J)=(1/2,0)$ channel below 
760 MeV, but the calculation with the full amplitudes\cite{NP} and 
the experimental data strongly suggest  the conclusion that the 
$\kappa(900)$ peak need not be the resonance. 
\item The $K\bar K$ correlation is attractive and so strong as to generate 
the bound state by $K\bar K$ loop alone in the isoscalar channel. 
\end{enumerate} 

It is impossible, however, to extract any information on the quark 
contents of the resonances from the analyses by the unitarized ChPT.  
Even if our conclusion on the mechanism of the generation of  
$f_0(980)$   and $a_0(980)$ is similar to 
Refs.\cite{WeinIsg,Julich1,Julich2}, 
we cannot say whether they are the $qq\bar q\bar q$ or the $K\bar K$   
molecule.  If the $f_0(980)$ appears the resonance but $a_0(980)$ the 
cusp by the hadronic dynamics, the $q\bar q$ scalar nonet should be 
attributed to the scalar resonances above 1 GeV. 
\\

Recently the complex pole search of the amplitudes on the unphysical 
sheets has been widely attempted in the study of the scalar 
meson spectroscopy.  We have also discussed the movement of the complex 
poles  of the $f_0(980)$ and $a_0(980)$ states .
The complex pole  is expected to be fundamental, parameterization- and 
process-independent, but the pole search is performed 
practically by use of a {\em model} amplitude, which is 
parameterized to {\em fit the data on the real axis  approximately}, 
and then the  pole position is the more model-dependent as it is the 
more distant from the real axis like the $\sigma$ and $\kappa$ 
cases\cite{Penning}.  

For example, it is shown that the amplitude coming from 
the $t-$ and $u-$channel $\rho$ 
meson exchange  in  the (0,0) channel develops the complex pole 
distant from the real axis at $(370-356{\rm i})$ MeV, for 
example\cite{ZouBugg}, 
while the phase shift stays at $50\sim 60^\circ$ 
at most\cite{Julich2,ZouBugg,LongLi}. A similar distant complex pole 
would be found in the (1/2,0) channel, because the role of the $\rho$ and 
$K^*$ exchange  and the behavior of the phase shift are quite similar 
to the ones in the (0,0) channel. Another example is the unitarized current 
algebra result\cite{CGL,JKH}, which gives a pole at 
$\sqrt{-i16\pi}f_\pi=463-i463$ MeV.  
An inverse example is the linear $\sigma$ model unitarized by the 
Pad\'e [1,1] approximation 
applied to the perturbation series up to the one-loop level\cite{BenLee}. 
In this model the physical $\sigma$ mass at which the phase shift 
crosses $90^\circ$ is set near the $\rho$ meson mass, but the 
complex poles 
in the unphysical sheet are found at near $(500- 300{\rm i})$ MeV. 
These values are close to the ones in the $\rho$ exchange model and the 
unitarized current algebra, but the physics on the real axis is quite 
different from each other.  The origin 
of the $\sigma$ pole is quite different from the one  of a pole near the real 
axis such as the $\rho$ meson pole. 
It would  be misleading, therefore, if we easily 
attribute any distant complex pole to the existence of an unstable 
meson state.   \\

There is an argument to relate the $t-$channel $\rho$ meson 
exchange  with the $s-$channel $\sigma$ resonance through the 
concept of the duality between the Regge pole exchange and the 
direct channel resonance\cite{Torn, Penning,Ochs}. 
This argument seems to be introduced so that the complex 
pole found in the $\rho$ exchange amplitude could be regarded as a 
root of $\sigma$ in the $s-$channel. If 
the $\pi^+\pi^-\to\pi^+\pi^-$ amplitude is expressed by the single 
Veneziano amplitude $V(s,t)$ in the Veneziano
model\cite{Veneziano},  the amplitude in the physical region near the 
pole at $s\sim M_\rho^2\sim M_\sigma^2$ is dominated not by 
the $t-$channel exchange term, $\sim g^2(s+t)/(M_\rho^2-t)$ but 
the $s-$channel pole term, $\sim g^2(s+t)/(M_\rho^2-s)$,  
and the both expressions cannot be added in order to avoid the double
counting.  Thus, it may be in trouble to add the $\rho$ meson 
exchange amplitude to the $\sigma$ pole term in the $s-$channel. 
Rather,the [1,1] Pad\'e approximant applied to the linear sigma model 
without the $\rho$ meson exchange\cite{BenLee} or the 
meson exchange model without the $s-$channel $\sigma$ pole\cite{Julich1} 
 would be consistent with the duality.

The duality would work severely for the models containing the explicit 
$\sigma$ field or the $\sigma$ pole besides the $\rho$ field in the theory.  
But the exchanged object in the above models is not any Regge pole but merely 
a simple pole, and  it is doubtful whether the complex pole 
found in the $\rho$ exchange amplitude represents really the 
$\sigma$ meson in the $s-$channel as discussed previously. If the model 
does not contain the $\sigma$ field as an independent degree of freedom, 
and if the mass enhancement peaked at 500 MeV, which may be called 
$\sigma$ state, comes from the two-pion correlation as 
the vacuum fluctuation owing to the spontaneous chiral symmetry 
 breaking, it would  need careful studies to understand  the duality 
 between the  Regge pole exchange and the $\sigma$ enhancement. \\

 We have argued the two-meson scattering processes within one of the 
unitarized chiral perturbation theories in this note and 
obtained rather unpopular conclusions on the scalar meson spectroscopy.   
There are many interesting 
phenomena such as the production and decay processes including the low 
mass scalar "mesons" in order to reveal the nature of them. The 
confirmation  of our conclusion and the consistency check with 
various processes should be pursued.

\hspace{3cm}

\end{document}